\documentclass[a4paper]{article}

\usepackage{a4wide}
\usepackage[UKenglish]{babel}
\usepackage{authblk}
\usepackage{graphicx,xcolor,colortbl}
	\graphicspath{{./}{figures/}}
\usepackage[colorlinks=true,linkcolor=blue,citecolor=red]{hyperref}
\usepackage{bbold,empheq}
\usepackage[inline]{enumitem}
\usepackage{subfigure}
\usepackage{xcolor}

\usepackage{pdfsync}

\usepackage{umoline}

\newcommand{\abs}[1]{\left\vert#1\right\vert}
\newcommand{\eps}{\varepsilon}
\newcommand{\n}{\mathbf{n}}
\newcommand{\R}{\mathbb{R}}

\allowdisplaybreaks

\begin{document}
\title{A sensitivity analysis of a mathematical model for the synergistic interplay of Amyloid beta and tau on the dynamics of Alzheimer's disease}

\author[1,2]{Michiel Bertsch}
\author[3]{Bruno Franchi}
\author[1]{Valentina Meschini}
\author[3]{Maria Carla Tesi}
\author[4]{Andrea Tosin}
\affil[1]{{\footnotesize Department of Mathematics, University of Roma ``Tor Vergata'', Roma, Italy}}
\affil[2]{{\footnotesize Istituto per le Applicazioni del Calcolo ``M. Picone'', Consiglio Nazionale delle Ricerche, Roma, Italy}}
\affil[3]{{\footnotesize Department of Mathematics, University of Bologna, Bologna, Italy}}
\affil[4]{{\footnotesize Department of Mathematical Sciences ``G. L. Lagrange'', Politecnico di Torino, Torino, Italy}}
\date{}

\maketitle

\begin{abstract}
We propose a mathematical model for the onset and progression of Alzheimer's disease based on transport and diffusion equations. We treat brain neurons as a continuous medium and structure them by their degree of malfunctioning. Three different mechanisms are assumed to be relevant for the temporal evolution of the disease:
\begin{enumerate*}[label=\roman*)]
\item diffusion and agglomeration of soluble Amyloid beta,
\item effects of phosphorylated  tau protein and
\item neuron-to-neuron prion-like transmission of the disease.
\end{enumerate*}
We model these processes by a system of Smoluchowski equations for the Amyloid beta concentration, an evolution equation for the dynamics of tau protein and a kinetic-type transport equation for the distribution function of the degree of malfunctioning of neurons. The latter equation contains an integral term describing the random onset of the disease as a jump process localized in particularly sensitive areas of the brain. We are particularly interested in investigating the effects of the synergistic interplay of Amyloid  beta and tau on the dynamics of Alzheimer's disease. The output of our numerical simulations, although in 2D with an over-simplified geometry, is in good qualitative agreement with clinical findings concerning both the disease distribution in the brain, which varies from early to advanced stages, and the effects of tau on the dynamics of the disease.

\medskip


\medskip

\end{abstract}

\section{Introduction}
Alzheimer's disease (AD) is a progressive neurodegenerative disease characterized by the formation of insoluble protein aggregates and loss of neurons and synapses. This causes a progressive decline in memory and other cognitive functions, and ultimately dementia. The processes leading to protein aggregation and neurodegeneration are only partially understood~\cite{AD}. AD was first described in 1907 by Alois Alzheimer who reported two pathological hallmarks in the brain: amyloid plaques in the extracellular medium and neurofibrillary tangles (NFTs). It was not until eight decades later that the major proteinaceous components of these lesions were identified. Amyloid plaques consist primarily of aggregates of Amyloid beta peptides (A$\beta$)~\cite{AD2}, whereas the main constituent of neurofibrillary tangles is  tau protein ($\tau$) in a hyperphosphorylated form~\cite{AD3}. Up to now, A$\beta$ and $\tau$ remain the major therapeutic targets for the treatment of AD. According to the 2019 World Alzheimer Report, it is estimated that there are presently 50 million people living with AD and related disorders, and this figure is expected to increase to 152 million by 2050 due to an increasingly aged population. As current treatments are purely for modest symptomatic relief, there is an urgent need for reliable and efficient computational models able to provide insights for effective therapies for the disease.

How A$\beta$ and $\tau$ interact to cause neurodegeneration remains a major knowledge gap in the field. There is substantial evidence that oligomeric A$\beta$ has a role in synapse degeneration, both in computational models and in human postmortem tissues, and that pathological forms of $\tau$ are sufficient to induce synapse loss and circuit dysfunction in models of tauopathy. In this context Jack e.a.\cite{jack_et_al_2} discuss a set of major biomarkers for AD and conclude that measures of both A$\beta$ and $\tau$ (both deposited and in CSF) are necessary to ascertain important features of AD pathology; this view stands in opposition to an amyloid-centric (or even taucentric) hypothesis. There is also increasing evidence that the progression of AD dementia is driven by the synergistic interaction between A$\beta$ and $\tau$, which is responsible for synaptic dysfunction, neurofibrillary tangle mediated neuron loss, and behavioral deficits~\cite{AD4,AD5,AD6}. For example, in~\cite{kara_et_al} the authors discuss possible relations between A$\beta$ aggregation, tau pathology and tau cell-to-cell transfer in AD; on the one hand the disease severity seems to correlate better with the severity of tau pathology than with that of A$\beta$ pathology, on the other hand A$\beta$ can enhance the formation of tau pathology~\cite{lewis_et_al,ittner_et_al} and cell-to-cell transfer of $\tau$~\cite{pooler_et_al}. 

Based on experiments with transgenic mice, in~\cite{ittner_et_al} the authors discuss several possible models for the interaction between A$\beta$ and $\tau$ as well as for their toxic effects. Basically, they discuss three different models:
\begin{enumerate*}[label=\alph*)]
\item A$\beta$ drives $\tau$ pathology by causing hyperphosphorylation of $\tau$, which in turn mediates toxicity in neurons;
\item $\tau$ mediates A$\beta$ toxicity and hence, A$\beta$ toxicity is critically dependent on the presence of $\tau$ for example, in the dendrite;
\item A$\beta$ and $\tau$ target cellular processed or organelles synergistically, thereby possibly amplifying each others toxic effects.
\end{enumerate*}

\subsection{The microtubule-associated~\texorpdfstring{$\boldsymbol{\tau}$}{}}
The $\tau$ protein is a microtubule-associated protein (MAP) that is encoded by the MAPT gene. The protein contains an amino-terminal projection domain, a proline-rich region, a carboxy-terminal domain with microtubule-binding repeats, and a short tail sequence. Tau has been reported to interact with many proteins, serving important scaffolding functions [16]. In particular, it acts in concert with heterodimers of $\alpha$- and $\beta$-tubulin to assemble microtubules and regulate motor-driven axonal transport. In the adult human brain  $\tau$ occurs in six isoforms, with either three or four microtubule-binding domains, which result from alternative splicing of exons 2, 3 and 10 of the MAPT gene. Tau is enriched inside the neurons. In nature neurons it is largely found in axons; in dendrites it is present in smaller amounts. The distinguishing factor that separates normal $\tau$ from that observed in patients with AD is its hyperphosphorylation. The longest isoforms of human $\tau$ contain 80 serine and threonine residues and five tyrosine residues, all of which can potentially be phosphorylated. In the disease state, the amount of hyperphosphorylated $\tau$ is at least three times higher than that in the normal brain~\cite{AD13}. It is not entirely clear, however, whether phosphorylation of $\tau$ at specific sites results in the pathogenicity observed in AD or whether it only requires a certain overall level of phosphorylation. Nevertheless, hyperphosphorylation of $\tau$ negatively regulates the binding of $\tau$ to microtubules, which compromises microtubule stabilization and axonal transport. Hyperphosphorylation also increases the capacity of $\tau$ to self-assemble and form aggregates from oligomers to fibrils, eventually leading to its deposition as NFTs . In addition, hyperphosphorylated $\tau$ has been shown to interfere with neuronal functions, causing reduced mitochondrial respiration, altered mitochondrial dynamics and impaired axonal transport. Tau pathology progresses through distinct neural networks, and in AD NFTs are prominent in the cortex in an early stage, and later appear in anatomically connected brain regions.  

Furthermore, it has been demonstrated that excess $\tau$ aggregates can be released into the extracellular medium, to be internalized by surrounding neurons and induce the fibrillization of endogenous $\tau$; this suggests a role for $\tau$ seeding in neurodegeneration~\cite{AD14}.

\subsection{Amyloid~\texorpdfstring{$\boldsymbol{\beta}$}{}}
\label{subsection Abeta}
Sequential cleavage of the amyloid precursor protein (APP) by $\beta$- and $\gamma$-secretase results in the generation of a range of A$\beta$ peptides from 39 to 43 amino acid residues in length, although A$\beta_{40}$ and A$\beta_{42}$ are the predominant species in vivo. The hydrophobic nature of the peptides, particularly A$\beta_{42}$ and A$\beta_{43}$, allows them to self-aggregate and form a myriad of species from dimers to small molecular weight oligomers, to protofibrils, to fibrils, ultimately leading to their deposition as amyloid plaques. Furthermore, A$\beta$ peptides can also undergo pyroglutamate modification at amino acid position three (A$\beta$3(pE)); this increases the stability, aggregation propensity and neurotoxicity compared to full-length, unmodified A$\beta$.  The mechanism by which excessive A$\beta$ accumulation occurs in sporadic AD remains unclear. Reduced A$\beta$ clearance or small increases in A$\beta$ production over a long period of time are potential mechanisms that result in the accumulation of A$\beta$ in the brain. The role of reduced A$\beta$ clearance is an important research topic which is widely discussed in the literature~\cite{Iliff_et_al,nedergaard_et_al}. The papers~\cite{Bordji_et_al,liu_et_al,Menendez_et_al} review various aspects of increased A$\beta$ production, a potentially intricate and complex process. Damaged neurons could cause increased production but $\tau$ may play a role in this process. If for example some A$\beta$ aggregates are formed and phosphorilated $\tau$ seeds are present in a proximal neuron, the aggregates may cause that $\tau$ relocates in the soma-dendritic compartment, where the errant $\tau$ can cause toxic effects to the neuron itself.

As we already pointed out, there are several possible models for the interaction between A$\beta$ and $\tau$. Some of them are presented in~\cite{ittner_et_al} In particular, the authors present what they call
\textit{$\tau$-axis hypothesis} of AD: progressively increasing levels of dendritic $\tau$ make neurons vulnerable to A$\beta$.

There is considerable debate regarding which of the A$\beta$ species is most neurotoxic. Increasing evidence suggests that small molecular weight oligomers correlate best with the disease and that insoluble amyloid plaques are not toxic~\cite{AD15}. It seems that the most toxic A$\beta$ is identified in A$\beta_{42}$.

We also stress that a longitudinal analysis of the main hallmarks of AD (though ignoring the role of $\tau$) is carried out in~\cite{BFMTT}. In particular, we see clearly in Figure 5 therein that the total amount
of soluble A$\beta$ in CSF decreases when the severity of the disease increases. This phenomenon is well known in clinical practice to establish an AD diagnosis (see e.g.~\cite{gabelli}).

\subsection{Aims of the model}
Increasing clinical evidence suggests that A$\beta$ and $\tau$ do not act in isolation and that there is significant crosstalk between these two molecules~\cite{AD12}. Experiments using primary neurons or neuronal cell lines have shown that application of A$\beta$ oligomers increases $\tau$ phosphorylation~\cite{AD10}. This suggests a link between A$\beta$ toxicity and $\tau$ pathology, but it is unclear how A$\beta$ and $\tau$ interact in pathological cases. Often $\tau$ is placed downstream of A$\beta$  in a pathocascade, providing support for the amyloid cascade hypothesis~\cite{AD11}. In this context we present a multiscale model for the onset and evolution of AD which accounts for the diffusion and agglomeration of A$\beta$ peptide, the protein $\tau$ and the spreading of the disease.

We stress that different spatial and temporal scales are needed to capture the complex dynamics of AD in a single model: microscopic spatial scales to describe the role of the neurons, macroscopic spatial and short temporal (minutes, hours) scales for the description of relevant diffusion processes in the brain, and large temporal scales (years, decades) for the description of the global evolution of AD.

Having these considerations in mind, the main purpose of the paper is to provide a reliable and flexible mathematical tool for better investigating mutual interaction mechanisms between A$\beta$ and $\tau$ and assess the resulting effect on the dynamics of Alzheimer's disease. Here reliability refers to the underlying mathematics, while flexibility is a crucial requirement which makes it possible to easily adapt the model to specific modelling hypotheses that one decides to test. In the case of AD flexibility becomes particularly important since, due to its extreme complexity, it is unthinkable to design a complete and global model from scratch.  
Any serious attempt in this direction would require an unrealistic amount of parameters, with values which are often unknown and difficult to estimate. Therefore we restrict ourselves to the design of the mathematical structure of the model, focussing on a limited amount of biomedical processes. In the present paper we focus on the role of possible interactions between A$\beta$ and phosphorylated $\tau$. In particular we test the hypothesis that their ``cooperation'' in the neuronal damage through different mechanisms is relevant for the evolution of the disease. In a simplified problem, which for example completely neglects the complex geometry of the brain, we show that in specific parameter regimes this is indeed the case.

\section{The mathematical model}
\label{MathModel}
The mathematical model employed in the present study derives from that extensively described, analyzed and motivated in~\cite{BFMTT,bertsch2017JPA,bertsch2018SIMA}. Modifications to the original model are made to cope with the main focus of the present study: the interplay between A$\beta$ and $\tau$.

We identify a portion of cerebral tissue by an either two- or three-dimensional bounded set $\Omega$. The space variable is denoted by $x$. Two time scales are needed to describe the evolution of the disease over a period of years: a short (i.e., rapid) $s$-scale, where the unit time coincides with hours, for the diffusion and agglomeration of A$\beta$~\cite{Meyer-Luhmann_nature}; and a long (i.e., slow) $t$-scale, where the unit time coincides with several months, for the progression of AD. Introducing a small constant $0<\eps\ll 1$, the relationship between these two time scales can be expressed as
\begin{equation}
	t=\eps s.
	\label{eq:ts}
\end{equation}

On the whole, the model that we propose for the present study reads as follows:
\begin{subequations}
	\begin{empheq}[left=\empheqlbrace]{align}
		&\partial_tf+\partial_a(fv[f])=J[f] & \text{in } \Omega\times [0,1]\times (0,T] \label{eq:system.f} \\
		&\eps\partial_tu_1-d_1\Delta{u_1}=-\alpha u_1U+\mathcal{F}[f]-\sigma_1u_1 & \text{in } Q_T=\Omega\times (0,T] \label{eq:system.u1} \\
		&\eps\partial_tu_2-d_2\Delta{u_2}=\frac{\alpha}{2}u_1^2-\alpha u_2U-\sigma_2u_2 & \text{in } Q_T \label{eq:system.um} \\
		&\eps\partial_tu_3=\frac{\alpha}{2}\sum_{3\leq j+k<6}u_ju_k & \text{in }Q_T \label{eq:system.uN} \\
		&\partial_t w=C_w(u_2-U_w)^{+}+\int_\Omega h_w(\abs{y-x})w(y,t)\,dy & \text{in } Q_T, \label{eq:system.w}
	\end{empheq}
	\label{complete_system2}
\end{subequations}
where in~\eqref{eq:system.u1},~\eqref{eq:system.um} we have set
$$ U:=\sum_{j=1}^{3}u_j $$
for conciseness.

We recall that, given a real number $z$, the symbol $z^+$ denotes the \textit{positive part} of $z$, i.e. $z^+=z$ if $z\geq 0$ and $z^+=0$ if $z<0.$

Equation~\eqref{eq:system.w} is the equation for the density $w(x,t)$ of intracellular phosphorylated $\tau$ at a point $x\in \Omega$ and time $t>0$. The equations~\eqref{eq:system.u1},~\eqref{eq:system.um},~\eqref{eq:system.uN} describe the dynamics of A$\beta$. Here 
\begin{enumerate}[label=(\roman*)]
\item $u_1(x,t)$ is the density of monomers in the point $x\in\Omega$ at time $t>0$;
\item $u_2(x,t)$ is the cumulative density of soluble oligomers, which are regarded collectively as a single compartment;
\item $u_3(x,t)$ is the density of senile plaques.
\end{enumerate}
In particular,~\eqref{eq:system.u1} and~\eqref{eq:system.um} are \textit{compartmental} Smoluchowski-type equations with diffusion, agglomeration and cleavage. A classical reference for Smoluchowski equations is~\cite{smoluchowski,drake}. Originally, these equations were introduced for the study of the aerosols; applications of Smoluchowski system to the description of the agglomeration of A$\beta$ amyloid appeared for the first time in~\cite{Murphy_Pallitto} and~\cite{AFMT}.

Such a compartmental model, which in particular does not distinguish the densities of the soluble oligomers based on their length, is justified by the fact that, according to the literature, there is no clinical evidence on the maximum length of toxic  A$\beta$ oligomers~\cite{hung2008}. Therefore, any precise value of such a length would be partly arbitrary. Although this may seem an over-simplification, in fact it is not because, as far as the results of the model are concerned, considering a more detailed description of the soluble oligomers with their precise lengths, such as e.g. in~\cite{BFMTT}, has proved not to add significant information~\cite{BFMPT,CTW}. The coefficient $\eps$ in front of the time derivatives is due to the relationship~\eqref{eq:ts}, in particular to the fact that, on the longer time scale $t$, the rate at which  agglomeration and diffusion of A$\beta$ take place is as high as $\frac{1}{\eps}$. Observe that we could rescale $\eps$, for example, by the proportion of an hour and a month: $\eps=0.0014 \times \tilde \eps$; this shows that $1/\eps$ is large but numerically quite manageable. Equation~\eqref{eq:system.uN} for the fibrils is written in the same spirit, except for the fact that fibrils are assumed not to move. Thus the equation for their concentration $u_3$ does not feature a diffusion term. We include in this compartment all the combinations of monomers and oligomers producing A$\beta$ entities other than those comprised in the compartments $1$ and $2$, taking further into account that senile plaques do not aggregate with each other. Finally, consistently with the compartmental nature of the model, we take the coagulation parameters constant and equal to \textcolor{red}{$\alpha>0$,} neglecting the fact that they may feature a dependence on the specific lengths of the aggregating oligomers, see~\cite{BFMTT}. We have chosen $\alpha$ as a positive parameter; one could also choose $\alpha$ age-dependent, or take $\alpha=0$ for young persons with healthy brains which do not produce toxic oligomers. Since senile plaques were also found in ``healthy brains'', we have preferred to take $\alpha>0$ and introduce in~\eqref{velocitytau} the modelling hypothesis that there is a positive threshold $\overline{U}$ for the density of oligomers below which toxic effects do not occur (see also the discussion following formula~\eqref{velocitytau}).

\label{pag:isotropic.D} The isotropic diffusion Laplace operator $\Delta$ in the Smoluchowski equations~\eqref{eq:system.u1} and~\eqref{eq:system.um} is a severe modelling simplification that completely ignores the complex structure of the cerebral parenchima, and in particular of the white matter. Soluble A$\beta$ diffuses in and is transported by cerebral liquid (essentially water) but in this paper we have chosen not to model the dynamics of the  cerebral fluid. Due to its flexibility the model could, at least in principle, be extended in this sense, but the topic is so complex that we prefer not to deal with it in the present paper. The brain is full of barriers (cell membranes, axons, myelin, etc.) which makes the microscopic (say at the size of an MRI-voxel), diffusion of the fluid highly anisotropic. For example, diffusion tends to be larger along the long axis of an axonal tract, and it could be influenced by myelin damage. An extensive discussion of the subject can be found e.g. in~\cite{aung_et_al,alexander_et_al,acosta_et_al}, in connection with the technique of Diffusion Tensor Imaging (DTI)  and its diagnostic clinical applications. Mathematically, good knowledge of the microscopic diffusion can be translated to macroscopic equations through homogenization techniques (see for example~\cite{FL1,FL2,FHL}, for homogenization procedures in the context of Smoluchowski equations applied to the modelization of AD). The resulting macroscopic and anisotropic diffusion operator is of the type $\nabla\cdot  (D^*(x,t)\nabla)$, where the diffusion matrix $D^*(x,t)$ keeps memory of the microscopic diffusion and the geometry of the brain. Obviously it is highly nontrivial to translate all this in quantitative information about $D^*(x,t)$. A second important problem is a possible convective contribution to the fluid flow. For example, the introduction of the concept of {\it glymphatic system} 
(see, e.g.~\cite{Iliff_et_al,nedergaard_et_al}) naturally leads to the modelling assumption that a convective contribution of the fluid flow is crucial for clearance mechanisms. For example, toxic A$\beta$ oligomers could be eliminated from the brain by the fluid flow. The precise mechanism of convective flow in the parenchima is still unknown, and its role in clearance mechanisms is still discussed in the medical literature.
We mention that an interesting new mathematical technique which gives quantitative information about the fluid flow in the brain was recently introduced in~\cite{koundal_et_al}.

Given the complexity of the fluid flow, in the present paper we have chosen to assume that the diffusion of A$\beta$ simply isotropic, while the terms $-\sigma_1u_1$ in~\eqref{eq:system.u1} and $-\sigma_2u_2$ in~\eqref{eq:system.um} are meant to keep into account several clearance phenomena, mainly due to phagocytic activity of the microglia. We are fully aware of the oversimplification of these choices, but we have preferred to focus on the possible interaction between A$\beta$ and $\tau$.

Equation~\eqref{eq:system.f} is a key ingredient of our approach of the model: a kinetic-type equation which describes the progression of the disease. Roughly speaking, $f=f(x,a,t)$ is the probability density of the degree of malfunctioning $a\in [0,1]$ of neurons located in $x\in\Omega$ at time $t>0$ and is such that $f(x,a,t)\,da$ represents the fraction of neurons in $x$ which at time $t$ have a degree of malfunctioning comprised between $a$ and $a+da$. For a precise mathematical formulation in terms of probability measures, see~\cite{bertsch2018SIMA}. We assume that $a$ close to $0$ stands for ``the neuron is healthy'' whereas $a$ close to $1$ stands for ``the neuron is dead''.

The progression of AD occurs on the longer time scale $t$, over decades, and is determined by the deterioration rate $v[f]=v[f](x,a,t)$, which we assume to have the following form:
\begin{equation}
	v[f](x,a,t)=C_\mathcal{G}\int_0^1(b-a)^+f(x,b,t)\,db+C_{\mathcal{S}}(1-a){\left(u_2(x,t)-\overline{U}\right)}^{+} +C_{\mathcal{W}}(1-a)w(x,t).
	\label{velocitytau}
\end{equation}
The integral term describes the propagation of AD among close neurons. The second term models instead the action of toxic A$\beta$ oligomers, leading ultimately to apoptosis. The threshold $\overline{U}>0$ indicates the minimal amount of toxic A$\beta$ needed to damage neurons. Finally, $C_\mathcal{G},C_\mathcal{S}>0$ are proportionality constants. The third term accounts for the toxicity of the phosphorylated $\tau$, whose density in the point $x\in\Omega$ at time $t>0$ is denoted by $w(x,t)$. Specifically, this term assumes that such a toxicity is proportional to the concentration $w$ through a proportionality constant $C_\mathcal{W}>0$ and that it is modulated by the current degree of malfunctioning of the neurons, in such a way that the more damaged the neurons the lower the impact of the toxic $\tau$ on them.

The term $J[f]=J[f](x,a,t)$ on the right-hand side of~\eqref{eq:system.f} describes the possible random onset of AD in portions of the domain $\Omega$ as a result of a microscopic stochastic jump process. The latter takes into account the possibility that the degree of malfunctioning of neurons randomly jumps to higher values due to external agents or genetic factors. The explicit expression of this term is
\begin{equation}
	J[f](x,a,t)=\eta\left(\int_0^1 P(t,x,a_\ast\to a)f(x,a_\ast,t)\,da_\ast-f(x,a,t)\right),
	\label{J formula}
\end{equation}
where $P(t,x,a_\ast\to a)$ denotes the probability that the degree of malfunctioning of neurons in the point $x\in\Omega$ jumps at time $t>0$ from $a_\ast$ to $a>a_\ast$. The coefficient $\eta>0$ is the jump rate.

It is worth stressing that~\eqref{eq:system.f}, together with the detailed expressions~\eqref{velocitytau},~\eqref{J formula} of the terms $v[f]$, $J[f]$, may be obtained from a mesoscopic description of a microscopic model of neuron-to-neuron interactions as shown in~\cite{bertsch2017JPA}.

As we have explained before, the main purpose of the paper is to introduce a simplified scheme of interaction between A$\beta$ and $\tau$, as proposed for instance in~\cite{ittner_et_al}. It is known that A$\beta$-plaques proximal to neuronal cell bodies can instigate, and exacerbate, $\tau$-pathology (\cite{pooler_et_al,selkoe_hardy,ittner_et_al}). However, there is no current evidence that $\tau$ influences A$\beta$-pathology in humans~\cite{selkoe_hardy}; moreover, a recent NIA-AA panel study suggests that all clinical diagnoses on the AD spectrum~\cite{NIAA-AA} should require the presence of A$\beta$-pathology biomarkers but not 
necessarily $\tau$ biomarkers. Based on the above we state an explicit modelling hypothesis used in our model: that {\it some minimal level of A$\beta$-aggregation is required to initiate $\tau$-pathology}. This assumption is encoded by the term $C_w(u_2-U_w)^{+}$ in~\eqref{eq:system.w}, where parameter $C_w>0$ is a proportionality constant. It is a built-in feature of our model that in equation~\eqref{eq:system.w} the evolution of $\tau$ phosphorilation is only due to the toxic effect of A$\beta$. This could suggest that the model is somehow too reminiscent of a ``purely amyloid hypothesis''. From the mathematical point of view, the addition of a source term in ~\eqref{eq:system.w} would not alter significantly the structure of the equations and, consequently, the outcomes of the simulations. Therefore we preferred this simplified formulation.

In~\eqref{velocitytau}, the formula of the deterioration rate $v[f]$, the term $C_{\mathcal{S}}(1-a){\left(u_2(x,t)-\overline{U}\right)}^{+}$ takes into account the toxic action of A$\beta$ and $C_{\mathcal{W}}(1-a)w(x,t)$ that of $\tau$.

The second term on the right-hand side of~\eqref{eq:system.w} describes the prion-like non-local spreading of the phosphorylated $\tau$ in possibly distant points of the brain according to the spatial kernel $h_w$. In Section~\ref{sect:mathins.tau} we will investigate in detail some properties of this integral term, which will further elucidate its physical meaning. We assume that the dynamics of  $\tau$ take place on the slow time scale $t$. In the absence of precise indications from the biomedical literature, this choice seems reasonable in view of the fact that  $\tau$ is especially involved in the progression of the disease rather than in the A$\beta$ agglomeration and the consequent formation of senile plaques.

To conclude the presentation of the model, we mention that the term $\mathcal{F}[f]=\mathcal{F}[f](x,a,t)$ in~\eqref{eq:system.u1} describes the production of A$\beta$ monomers by neurons, taking into account that, up to a certain extent, damaged neurons increase such a production. In view of these considerations, we choose
\begin{equation}
	\mathcal{F}[f](x,t)=C_{\mathcal{F}}\int_0^1(\mu_0+a)(1-a)f(x,a,t)\,da.
	\label{eq:F}
\end{equation}
Here, the small constant $\mu_0>0$ accounts for A$\beta$ production by healthy neurons while the factor $1-a$ expresses the fact that dead neurons do not produce amyloid. As usual, $C_\mathcal{F}>0$ is a proportionality constant. Again the flexibility of the model makes it possible to take into account the possible role of $\tau$ in the production of  A$\beta$ monomers, as we have sketched in Section~\ref{subsection Abeta}.

\subsection{Initial and boundary conditions}
\label{modeltau}
As far as boundary conditions are concerned, we assume that $\partial\Omega$ consists of two smooth disjoint parts, say $\partial\Omega_0$ and $\partial\Omega_1$, being $\partial\Omega_0$ the outer boundary which delimits the considered portion of cerebral tissue and $\partial\Omega_1$ the inner boundary of the cerebral ventricles. On $\partial\Omega_0$ we prescribe classical no-flux conditions for all the concentrations. Conversely, on $\partial\Omega_1$ we prescribe a Robin condition for the concentrations of the A$\beta$ oligomer mimicking their removal by the cerebrospinal fluid through the choroid plexus~\cite{Iliff_et_al,serot_et_al}. On the whole, we have then:
\begin{equation}
	\begin{cases}
		\nabla{u_i}\cdot\n=0 & \text{on } \partial\Omega_0,\ i=1,2 \\
		\nabla{u_i}\cdot\n=-\beta u_i & \text{on } \partial\Omega_1,\ i=1,2 \\
		\nabla{w}\cdot\n=0 & \text{on } \partial\Omega,
	\end{cases}
	\label{eq:BC}
\end{equation}
where $\beta>0$ is a proportionality parameter and $\n$ the outward normal unit vector to $\partial\Omega$. 
The choice of the right-hand side of the above Robin condition, in absence of experimental data, is the simplest
possible. Maybe a more realistic choice would be
$$ \nabla{u_i}\cdot\n=-h(u_i) \quad  \text{on } \partial\Omega_1,\ i=1,2, $$
where $h(u_i)\approx\beta u_i$ as $u_i\to 0$, and $h(u_i)\approx 1$ as $u_i\to\infty$.
 
We also complement system~\eqref{complete_system2} with a proper set of initial conditions:
\begin{equation}
	f(x,a,0)=f_0(x,a), \qquad u_i(x,0)=u_{0,i}(x)\ (i=1,2,3), \qquad w(x,0)=0.
	\label{eq:IC}
\end{equation}

A numerical discretisation of the initial/boundary-valued problem~\eqref{complete_system2}-\eqref{eq:BC}-\eqref{eq:IC} can be set up straightforwardly by adapting the one described in detail in~\cite{BFMTT}.

\subsection{Mathematical insights from the model}
\label{sect:mathins.tau}
In order to familiarise with model~\eqref{complete_system2} and to grasp the biological meaning of several complex terms appearing in the equations, in this section we address some biologically relevant issues that can be tackled with suitably simplified but illustrative instances of~\eqref{complete_system2}. These case-dependent simplifications will make~\eqref{complete_system2} more amenable to analytical investigations and, at the same time, will provide a more intuitive feeling about the whole model and the way in which it works.

\subsubsection{AD onset}
\label{sect:AD_onset}
The first issue that we investigate by means of~\eqref{complete_system2} is how quickly completely healthy neurons at a point $x\in\Omega$ of the brain tissue start deteriorating. To dig into this aspect, we mainly consider equation~\eqref{eq:system.f} together with the expression~\eqref{velocitytau} of the deterioration rate of the neurons and we prescribe the initial conditions~\eqref{eq:IC} with in particular
\begin{equation}
	f_0(x,a)=g(x)\delta(a), \quad u_{2,0}(x)=0,
	\label{eq:onset.IC}
\end{equation}
$g$ being any spatial distribution of the neurons. These conditions depict an initially healthy brain.

Next, we assume that up to a certain time $t^\ast>0$ we have $u_2(x,t)<\overline{U},\,U_w$ for all $x\in\Omega$, so that for $t\leq t^\ast$ it results $(u_2(x,t)-U_w)^+=0$ in~\eqref{eq:system.w} and analogously $(u_2(x,t)-\overline{U})^+=0$ in~\eqref{velocitytau}. Consequently, from~\eqref{eq:system.w} we obtain $w(x,t)=0$ and from~\eqref{velocitytau}
$$ v[f](x,a,t)=C_\mathcal{G}\int_0^1(b-a)^+f(x,b,t)\,db, $$
so that~\eqref{eq:system.f} takes the form
\begin{equation}
	\partial_tf(x,a,t)+C_\mathcal{G}\partial_a\left(f(x,a,t)\int_0^1\int_0^1(b-a)^+f(x,b,t)\,db\right)=J[f](x,a,t),
	\label{eq:onset.f}
\end{equation}
where the term $J[f]$ is given by~\eqref{J formula}. We may now take advantage of~\eqref{eq:onset.f} to investigate the local onset of AD in an initially healthy brain.

First of all, we notice that $v[f](x,1,t)=0$ for all $x\in\Omega$ and $t>0$. If we prescribe the further boundary condition $f(x,0,t)=0$ for all $x\in\Omega$ and $t>0$, which is consistent with the fact that no degree of malfunctioning $a<0$ is admitted by the model, then by integrating~\eqref{eq:onset.f} with respect to $a$ and noticing from~\eqref{J formula} that $\int_0^1J[f](x,a,t)\,da=0$ by construction, we discover
$$ \partial_t\rho(x,t)=0, $$
where
$$ \rho(x,t):=\int_0^1f(x,a,t)\,da $$
denotes the local percentage of neuronal mass at time $t$. Therefore the mass of neurons is conserved both locally and globally, consistently with the modelling assumption that neurons are fixed in space and are neither created nor destroyed. It is worth pointing out that in our model the possible brain atrophy caused by AD is indeed not modelled explicitly as a physical loss of neuronal mass.

Second, multiplying~\eqref{eq:onset.f} by $a$ and then integrating as before we obtain an evolution equation for the local mean degree of neuronal malfunctioning
$$ A(x,t):=\frac{1}{\rho(x)}\int_0^1af(x,a,t)\,da, $$
where $\rho(x)$ stands for the constant-in-time local percentage of neuronal mass at $x\in\Omega$. Specifically:
\begin{equation}
	\partial_tA(x,t)-\frac{C_\mathcal{G}}{\rho(x)}\int_0^1\int_0^1(b-a)^+f(x,b,t)f(x,a,t)\,db\,da=\frac{1}{\rho(x)}\int_0^1aJ[f](x,a,t)\,da.
	\label{eq:onset.A}
\end{equation}
The quantity $A(x,t)$ provides an observable macroscopic picture of AD trends in the brain, in particular the initial onset for small $t$. We anticipate that in the numerical simulations of Section~\ref{sec:num_res} we will explore instead its large time trend.

To make the term $\int_0^1aJ[f](x,a,t)\,da$ amenable to explicit computations, we take for the jump probability $P(t,x,a_\ast\to a)$, cf.~\eqref{J formula}, the representative expression already introduced in~\cite{BFMTT,bertsch2017JPA,bertsch2018SIMA}:\footnote{\label{foot:chi}Here and henceforth, $\chi_E$ denotes the characteristic function of the set $E$, i.e.
$$ \chi_E(a)=
	\begin{cases}
		1 & \text{if } a\in E \\
		0 & \text{otherwise}	.
	\end{cases} $$}
$$ P(t,x,a_\ast\to a)=\frac{2}{1-a_\ast}\chi_{[a_\ast,\,\frac{1+a_\ast}{2}]}(a), $$
namely a uniform probability distribution of the post-jump degree of malfunctioning $a$ in the interval $[a_\ast,\,\frac{1+a_\ast}{2}]$. This complies with the requirement of physical consistency $a_\ast\leq a\leq 1$ for all pre-jump degree of malfunctioning $a_\ast\in [0,\,1]$. Notice that, for the sake of simplicity, such a $P$ is homogeneous in time and space. With this choice, we get
\begin{equation}
	\frac{1}{\rho(x)}\int_0^1aJ[f](x,a,t)\,da=\frac{\eta}{2}(1-A(x,t)).
	\label{eq:onset.aJ}
\end{equation}

For $a,\,b\in [0,\,1]$ it can be checked that $0\leq (b-a)^+\leq b(1-a)$, whence a simple computation yields
\begin{equation}
	0\leq\frac{C_\mathcal{G}}{\rho(x)}\int_0^1\int_0^1(b-a)^+f(x,b,t)f(x,a,t)\,db\,da\leq C_\mathcal{G}\rho(x)A(x,t)(1-A(x,t)).
	\label{eq:onset.(b-a)^+}
\end{equation}

Plugging~\eqref{eq:onset.aJ},~\eqref{eq:onset.(b-a)^+} into~\eqref{eq:onset.A} we obtain the estimates
$$ \frac{\eta}{2}(1-A(x,t))\leq\partial_tA(x,t)\leq C_\mathcal{G}\rho(x)A(x,t)(1-A(x,t))+\frac{\eta}{2}(1-A(x,t)), $$
which, by an integration in time from the initial condition $A(x,0)=0$ derived from~\eqref{eq:onset.IC}, produce
\begin{equation}
	1-e^{-\frac{\eta}{2}t}\leq A(x,t)\leq 1-\frac{C_\mathcal{G}\rho(x)+\frac{\eta}{2}}{C_\mathcal{G}\rho(x)+\frac{\eta}{2}e^{(C_\mathcal{G}\rho(x)+\frac{\eta}{2})t}}.
	\label{eq:onset.A_bounds}
\end{equation}

For $t\to 0^+$ both bounds are asymptotic to $\frac{\eta}{2}t$, therefore for $t$ small we infer
$$ A(x,t)\sim\frac{\eta}{2}t. $$
This indicates that the rate of AD onset is dictated by the frequency $\eta$ at which random neuron damages appear in the brain. Notice that if $\eta=0$ we obtain the exact solution $A\equiv 0$, which confirms that random jumps are essential in the model to trigger the AD onset in a completely healthy brain. On the other hand,~\eqref{eq:onset.A_bounds} reveals that for larger times, however such that the A$\beta$-related terms in~\eqref{eq:system.um},~\eqref{eq:system.w} are still deactivated, the subsequent rate of AD progression depends also on the parameter $C_\mathcal{G}$, which tunes the strength of AD propagation among close neurons, and on the local percentage of neuronal mass $\rho(x)$.

\subsubsection{Effect of the non-local term in~\texorpdfstring{\eqref{eq:system.w}}{}}
Here we explore instead the effect of the non-local term
\begin{equation}
	I(x,t):=\int_\Omega h_w(\abs{y-x})w(y,t)\,dy
	\label{eq:I(x,t)}
\end{equation}
in~\eqref{eq:system.w} with $\Omega\subset\R^n$ ($n=2,3$ from the physical point of view). Without loss of generality, we assume that the function $h_w:\R_+\to\R_+$ is such that $h_w'(0^+)=0$. Furthermore, we assume that $h_w$ is compactly supported in $[0,R]\subset\R$, where $R>0$ denotes the maximum distance at which the prion-like influence of the phosphorylated $\tau$ is supposed to be effective. Hence
$$ I(x,t)=\int_{B_R(x)}h_w(\abs{y-x})w(y,t)\,dy. $$

Assume that $R$ is sufficiently small, so that $y$ is close to $x$, and that the functions $h_w$, $w$ are smooth enough about $0^+$ and $x$, respectively. By Taylor expansion we find:
\begin{align*}
	h_w(\abs{y-x}) &= h_w(0^+)+\frac{1}{2}h_w''(0^+)\abs{y-x}^2+o(\abs{y-x}^2) \\
	w(y,t) &= w(x,t)+\sum_{h=1}^{n}\partial_{x_h}w(x,t)(y_h-x_h) \\
	&\phantom{=} +\frac{1}{2}\sum_{h,k=1}^{n}\partial^2_{x_hx_k}w(x,t)(y_h-x_h)(y_k-x_k)+o(\abs{y-x}^2),
\end{align*}
thus
\begin{align*}
	h_w(\abs{y-x})w(y,t) &= h_w(0^+)w(x,t)+h_w(0^+)\sum_{h=1}^{n}\partial_{x_h}w(x,t)(y_h-x_h) \\
	&\phantom{=} +\frac{1}{2}h_w(0^+)\sum_{h,k=1}^{n}\partial^2_{x_hx_k}w(x,t)(y_h-x_h)(y_k-x_k) \\
	&\phantom{=} +\frac{1}{2}h_w''(0^+)w(x,t)\abs{y-x}^2+o(\abs{y-x}^2).
\end{align*}
Neglecting the remainder $o(\abs{y-x}^2)$, we can therefore approximate $I(x,t)$ locally as
\begin{align*}
	I(x,t) &\approx h_w(0^+)w(x,t)\int_{B_R(x)}dy
		+h_w(0^+)\sum_{h=1}^{n}\partial_{x_h}w(x,t)\int_{B_R(x)}(y_h-x_h)\,dy \\
	&\phantom{\approx} +\frac{1}{2}h_w(0^+)\sum_{h,k=1}^{n}\partial^2_{x_hx_k}w(x,t)\int_{B_R(x)}(y_h-x_h)(y_k-x_k)\,dy \\
	&\phantom{\approx} +\frac{1}{2}h_w''(0^+)w(x,t)\int_{B_R(x)}\abs{y-x}^2\,dy.
\end{align*}

Let $\omega_n$ denote the volume of the unit ball in $\R^n$. By switching to polar coordinates, we find
\begin{align*}
	& \int_{B_R(x)}dy=\omega_nR^n, \\
	& \int_{B_R(x)}(y_h-x_h)\,dy=0, \\
	& \int_{B_R(x)}(y_h-x_h)(y_k-x_k)\,dy=
	\begin{cases}
		\dfrac{\omega_n}{n+2}R^{n+2} & \text{if } h=k \\[3mm]
		0 & \text{if } h\neq k,
	\end{cases} \\
	& \int_{B_R(x)}\abs{y-x}^2\,dy=\frac{n\omega_n}{n+2}R^{n+2},
\end{align*}
whence finally
$$ I(x,t)\approx\omega_nR^n\left(h_w(0^+)+\frac{n}{2(n+2)}h_w''(0^+)\right)w(x,t)+\frac{\omega_nR^{n+2}}{2(n+2)}h_w(0^+)\Delta{w(x,t)}. $$

This shows that the local contribution of $I(x,t)$ to the spreading of the phosphorylated $\tau$ consists in a source term proportional to the quantity of $\tau$ already present in the site $x$ and in a linear diffusion.

The local approximation of $I(x,t)$ elucidates also the role of the pointwise values of $h_w$, $h_w''$ in $r=0$. If, for instance, we take \footnote{For the meaning of $\chi_{[0,\,R]}(r)$, cf. footnote~\ref{foot:chi} on page~\pageref{foot:chi}.}
$$ h_w(r)=\frac{1}{\omega_nR^n}\chi_{[0,R]}(r), $$
which makes $I(x,t)$ a uniform average of $w$ in a neighbourhood of $x$ of radius $R$, then $h_w(0^+)=\frac{1}{\omega_nR^n}$ and $h_w''(0^+)=0$, whence
\begin{equation}
	I(x,t)\approx w(x,t)+\frac{R^2}{2(n+2)}\Delta{w(x,t)}. 
	\label{eq:I.approx} 
\end{equation}
This implies that, at the leading order in $R$, $I$ consists in a pointwise source and, at higher order, an isotropic diffusion term responsible for the spatial spreading of the phosphorylated $\tau$.
Notice that such a pointwise approximation of $I(x,\,t)$ is consistent with the isotropic approximation of the diffusion tensor in~\eqref{eq:system.u1},~\eqref{eq:system.um} discussed on page~\pageref{pag:isotropic.D}.

\subsubsection{A\texorpdfstring{$\boldsymbol{\beta}$}{}-distribution in a healthy brain}
Determining the distribution of A$\beta$ in a healthy brain is relevant because it provides a consistent initial condition for studying AD initiation processes. Moreover, it may also suggest how the geometry of the domain $\Omega$, thus the brain conformation, affects the result.

We already discussed a proper set of initial conditions for model~\eqref{complete_system2} simulating a healthy brain, cf.~\eqref{eq:IC},~\eqref{eq:onset.IC}. The point here is to characterise the resulting distribution $u_1$ of non-toxic A$\beta$ monomers, which is essentially ruled by~\eqref{eq:system.u1}.

Assume that the individual is young, say under $40$ years of age, and healthy, in particular with no genetic predisposition to AD. Then it makes sense to consider $\alpha=0$ in~\eqref{complete_system2} and $\eta=0$ in~\eqref{J formula}, which imply no coagulation of non-toxic A$\beta$ monomers into toxic polymers and no random damages of the neurons, respectively. From Section~\ref{sect:AD_onset} we know that the latter condition produces $A\equiv 0$, which in turn implies that the distribution $f$ of the degree of malfunctioning has remained proportional to a Dirac delta centred in $a=0$ like in the initial condition~\eqref{eq:onset.IC}. This is so because $A$ is, by definition, the (local) mean of $a\in [0,\,1]$, cf.~\eqref{eq:onset.A}, and if the mean is zero then the probability distribution of $a$ has to be concentrated at $0$. Hence $f(x,a,t)=g(x)\delta(a)$.

On the whole, from~\eqref{eq:F} we obtain
$$ \mathcal{F}[f](x,t)=C_\mathcal{F}\mu_0g(x), $$
which plugged into~\eqref{eq:system.u1} yields
$$ \varepsilon\partial_tu_1-d_1\Delta{u_1}=C_\mathcal{F}\mu_0g(x)-\sigma_1u_1. $$
Since $\varepsilon$ is small, on the time scale $t$ of the AD progression we may well assume that the distribution of A$\beta$ monomers is quasi-static. Consequently, recalling also the boundary conditions~\eqref{eq:BC}, the physiological distribution $u_1=u_1(x)$ of A$\beta$ monomers in a healthy brain is fully characterised by the following reaction-diffusion problem:
\begin{equation}
	\begin{cases}
		-d_1\Delta{u_1}+\sigma_1u_1=C_\mathcal{F}\mu_0g(x) & \text{in } \Omega \\
		\nabla{u_1}\cdot\n=0 & \text{on } \partial\Omega_0 \\
		\nabla{u_1}\cdot\n=-\beta u_1 & \text{on } \partial\Omega_1.
	\end{cases}
	\label{eq:u1.healthy}
\end{equation}

As already recalled in Section~\ref{sect:AD_onset}, the function $g$ describes the concentration (statistical distribution) of the neurons in the cerebral tissue. The simplest choice is to consider a homogeneous distribution, which amounts to letting\footnote{For the meaning of $\chi_\Omega(x)$, cf. footnote~\ref{foot:chi} on page~\pageref{foot:chi}.}
$$ g(x)=\frac{1}{\abs{\Omega}}\chi_\Omega(x), $$
where $\abs{\Omega}$ denotes the Lebesgue measure (the area or the volume in the physically relevant two- or three-dimensional cases) of $\Omega$. Non-homogeneous distributions are obviously also possible, if one wants to take into account the heterogeneity of the neuron concentration in the brain. Problem~\eqref{eq:u1.healthy} clearly shows that such a heterogeneity in general has an impact on the physiological distribution of A$\beta$ monomers.

\section{Numerical results}
\label{sec:num_res}
The main purpose of this section is to numerically validate the model previously introduced. More specifically, our model can be schematized as follows:
\begin{itemize}
\item the A$\beta$ diffuses in the cerebral parenchima undergoing a coagulation process leading ultimately to the formation of senile plaques. In its oligomeric form, the A$\beta$ induces the phosphorylation of
the intra-neural $\tau$ protein.
\item A$\beta$ and $\tau$ have a synergistic toxic effect on the neurons.
\end{itemize}
The model we present here, for what purely concerns the A$\beta$, is akin to the model presented in~\cite{BFMTT} within a strict Amyloid Cascade Hypothesis, using a compartmental model with only two aggregating species monomers and dimers. Oligomers with length longer than two are considered as plaques, thus assuming $N=3$.

In Section~\ref{MathModel} we extensively discussed various other modelling simplifications (such as that of isotropic diffusion). As we have announced in the Introduction, in the present section we add two more essential simplifications: we present two-dimensional numerical simulations and we neglect the geometric complexity of the brain. In fact, in all figures the numerical domain is a rectangle and represents a two dimensional section of the brain where the upper part is the frontal region, the lower part is the occipital region and the two inner rectangles are the ventricles.

Finally, the non-local term $I(x,t)$, see~\eqref{eq:I(x,t)}, is replaced in~\eqref{eq:system.w} by its leading term $w(x,t)$, cf.~\eqref{eq:I.approx}.

The model contains many parameters which are often not known. We fix some of them (see Table~\ref{table1}) and vary other ones (see Table~\ref{table2}), and we have done this in such a way that the numerical experiments give some first insights concerning possible synergistic effects of $\tau$ and A$\beta$. More precisely, we focus on the following topics:
\begin{itemize}
\item neurodegeneration in absence of $\tau$ (see Section~\ref{no tau});
\item the effects of including $\tau$ in the model and its synergistic interaction with A$\beta$ (see Section~\ref{with tau});
\item the impact of the agglomeration rate $\alpha$ on the disease progression (see Section~\ref{sect alpha});
\item the effects of the removal rate $\beta$ in presence of $\tau$ (see Section~\ref{sect beta});
\item the effects of the threshold value $\overline{U}$ (see Section~\ref{no tau}).
\end{itemize} 
 
These target topics are discussed in separate paragraphs in this section  presenting and discussing the numerical results of each dedicated simulation of the mathematical model for Alzheimer's disease described in Section~\ref{MathModel}. We show the evolution of AD in two different modeling schemes: the first does not take into account the role of  $\tau$, whereas the second includes  $\tau$. We show spatial plots of the evolution of the  degree of neuronal dysfunction, expressed by the quantity $f$ in~\eqref{eq:system.f}, as well as spatial plots of the concentration of the toxic polymers, plaques and (phosphorylated) $\tau$.

As a consequence of the compartmental description of the A$\beta$ oligomers, we find that the numerical results are quite sensitive to the constant $\overline{U}$ in~\eqref{velocitytau}, which we will therefore tune accurately so as to observe in practice an evolution of the disease. In particular, a too high value of $\overline{U}$ may easily hinder the effective toxicity of the cumulative density $u_2$ of A$\beta$, thereby preventing the occurrence of the dynamics of AD. As a matter of fact, since the phosphorylated $\tau$ contributes to the overall toxicity~\cite{AD5}, the spread of the disease may also occur for a relatively high value of $\overline{U}$.

Besides $\overline{U}$, the equation for $\tau$ (see~\eqref{eq:system.w}) contains another threshold constant $U_w$ to which our model results very sensitive. There exists a small critical range around $U_w=0.01$: for smaller values of $U_w$ the role of A$\beta$ is reduced to the mere phosphorylation of tau and significant deterioration is due to tau only; conversely, for higher values of $U_w$ not enough tau is phosphorylated to play a significant role in the deterioration process. Therefore we fix $U_w=0.01$ in all numerical tests, a value for which there is a synergistic effect of tau and A$\beta$ in the deterioration process.  

This choice is consistent with the model we are developing, which is supported by patient biomarkers~\cite{AD10} and experiments on transgenic mice~\cite{ittner_et_al}.

In both simulation set-ups (with and without $\tau$) we carry out a sort of sensitivity analysis of the model outcomes with respect to the parameters $\alpha$, $\beta$ and $\overline{U}$, also in view of possible clinical implications.

First of all, we consider the parameter $\beta$, which enters the model through condition~\eqref{eq:BC} at the boundary of the cerebral ventricles. Small values of $\beta$ mean that a small amount of A$\beta$ is removed from the CSF through the choroid plexus. To test the sensitivity of the outcomes, we take  for $\beta$ two values: $\beta= 0.01$ and $\beta =1$. Another significant parameter is $\alpha$ (the coagulation parameter in~\eqref{eq:system.u1},~\eqref{eq:system.um}).  We choose two different orders of magnitude of $\alpha$ to observe how it affects the dynamics of A$\beta$ in the brain. We run a simulations campaign (both with and without including $\tau$) varying $\alpha$, $\beta$ and $\overline{U}$ but keeping all other parameters fixed (see the captions in the sequel), with the values in Table~\ref{table1}. These values represent orders of magnitude and have a proper unit measure that can be deduced from the equations in which they are involved.

\renewcommand{\arraystretch}{1.5}
\begin{table}[!t]
\caption{Values of the fixed parameters}
\begin{center}
\begin{tabular}{lcccccccc}
\hline
\rowcolor{lightgray} $C_G$&$C_S$ & $D$ & $\overline{\sigma}$ & $\mu_0$ & $C_F$ & $C_W$& $\overline{U}_w$ \\
\hline
\hline
$10^{-1}$ & $10^{-3}$ & $10^{-2}$ & $1$ & $10^{-2}$ & $10$ & $10$ & $10^{-2} $\\
\hline
\end{tabular}
\end{center}
\label{table1}
\end{table}

We perform numerical tests of the various choices of the parameters presented in Table~\ref{table2}. The insights obtained from these simulations, especially those related to possible clinical treatments and procedures will be discussed in section~\ref{Discussion}. 

\renewcommand{\arraystretch}{1.5}
\begin{table}[!t]
\caption{All the simulated cases and involved parameters with respective values}
\begin{center}
\begin{tabular}{lcccc}
\hline
& Inclusion of $\tau$ & $\alpha$ & $\beta$ & $\overline{U}$ \\
\hline
\hline
\rowcolor{lightgray} Case 1 & No & $10$ & $1$ & $10^{-2}$ \\
\hline
Case 1.1 & No & $10$ & $1$ & $10^{-1}$ \\
\hline
\hline
\rowcolor{lightgray} Case 2 & Yes & $10$ & $1$ & $10^{-2}$ \\
\hline
Case 2.1 & Yes & $1$ & $1$ & $10^{-2}$ \\
\hline
Case 2.2 & Yes & $10$ & $10^{-2}$ & $10^{-2}$ \\
\hline
\end{tabular}
\end{center}
\label{table2}
\end{table}

\subsection{Neurodegeneration in absence of~\texorpdfstring{$\boldsymbol{\tau}$}{}}
\label{no tau}
\begin{figure}[!t]
\centering
\includegraphics[width=0.9\textwidth,keepaspectratio]{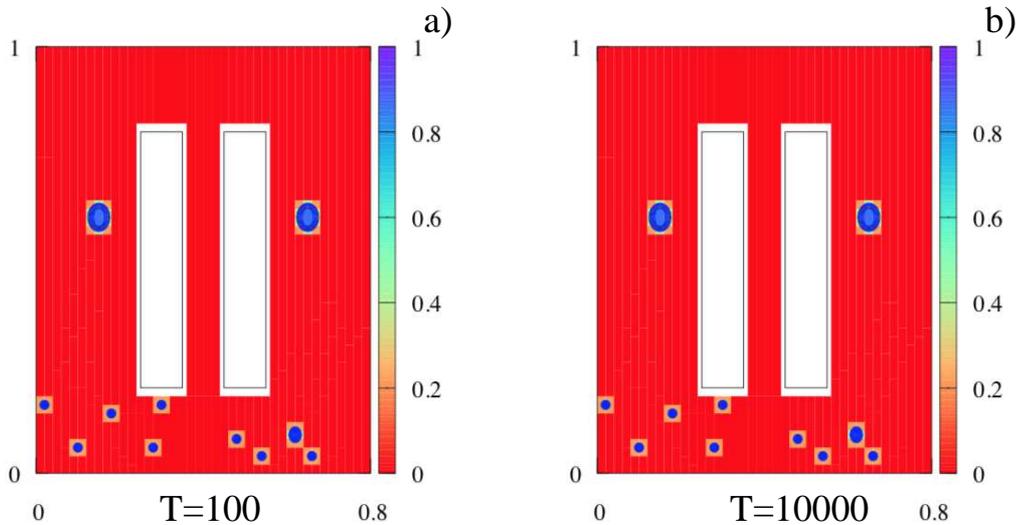}
\caption{(Case 1.1) spatial plots of the degree of malfunctioning of the brain for $\alpha=10$,  $\beta=1$, $\overline{U}=0.1$ and without $\tau$ at two time instants: a) $T=100$, b) $T=10000$.}
\label{Figure1_1}
\end{figure}

In this subsection we simulate the model for Alzheimer's disease which was already presented in~\cite{BFMTT}, without the presence of $\tau$ but with $N=3$. The only toxic oligomers are dimers and therefore the overall toxicity level is much lower than in~\cite{BFMTT}, in which $N=50$. The toxicity level enters the formula for the degree of malfunctioning of the brain, which provides information on the health state of the brain.
The parameter $\alpha$, which regulates the coagulation of monomers and dimers, is taken equal to $10$, while $\beta$, which refers to the portion of toxic oligomers extracted from the brain, is equal to $1$. Given these parameter values, we choose the threshold value $\overline{U}$ for toxic $u_2$ equal to $10^{-2}$, which turns out to be the minimal order of magnitude to allow the disease to take place. In fact, if we set $\overline{U}=0.1$ the degree of malfunctioning of the brain does not evolve in time and the disease does not spread (see Figure~\ref{Figure1_1}).

\begin{figure}[!t]
\centering
\includegraphics[width=0.9\textwidth,keepaspectratio]{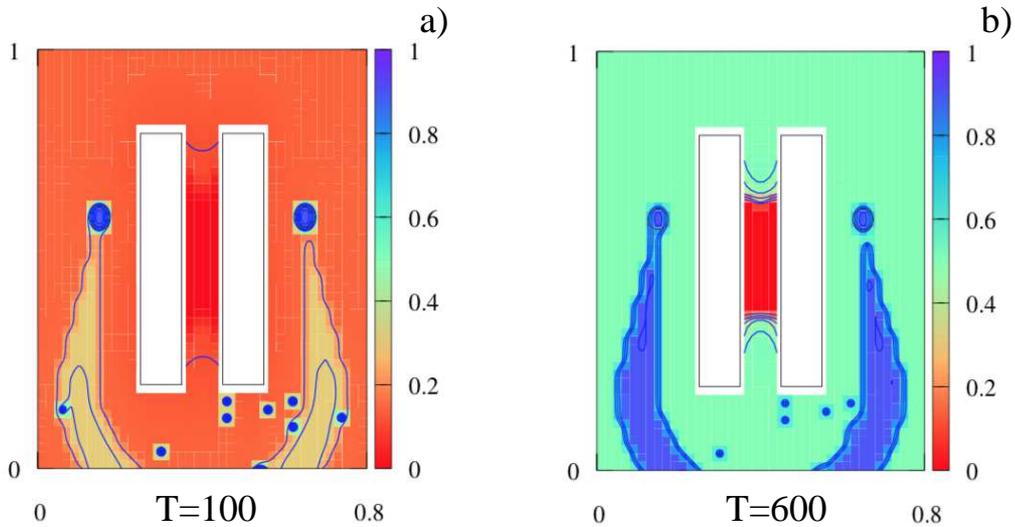}
\caption{(Case 1) spatial plots of the degree of malfunctioning of the brain for $\alpha=10$, $\beta=1$, $\overline{U}=10^{-2}$ and without $\tau$ at two time instants: a) $T=100$, b) $T=600$}
\label{Figure1}
\end{figure}

The numerical outputs with $\overline{U}=10^{-2}$ are shown in Figure~\ref{Figure1}, where the spatial plots of the degree of neuronal malfunctioning $f$ at different times are indicated. The disease originates from some random sources (the blue spots) and does not evolve. Due to the  propagation of synaptic dysfunction, it spreads to other portions of the brain until it reaches a numerical equilibrium configuration, in the sense that 
the spatial plot of $f$ does not change anymore if time evolves. In this ``quasi-steady state''\footnote{The concept of quasi-steady state refers to a very slow evolution which, in the mathematical literature, is often referred to as quasi-static.} we can see that there is a blue zone (dead part) but the brain is not completely ill. This result is clearly different from that obtained in~\cite{BFMTT}, in which the whole brain is blue thus meaning basically that the patient is dead.

The interpretation of this new phenomenon is the following, Since the only toxic oligomers are dimers and having a coagulation factor $\alpha=10$, the A$\beta$ monomers which aggregate to form dimers remain only briefly in this toxic state and quickly become plaques which are non toxic. Therefore the toxicity of the whole system is much lower. The formation of the quasi steady state has a similar explanation: at a certain point the concentration of toxic oligomers $u_2$ goes below the threshold level $\overline{U}$, because toxic dimers quickly agglomerate into plaques, and the degree of malfunctioning does not evolve any further. In~\cite{BFMTT} the phenomenon was not observed since there $N=50$ instead of $N=3$.

\subsection{The interaction between A\texorpdfstring{$\boldsymbol{\beta}$}{} and~\texorpdfstring{$\boldsymbol{\tau}$}{} and the role of~\texorpdfstring{$\boldsymbol{\tau}$}{}}
\label{with tau}
\begin{figure}[!t]
\centering
\includegraphics[width=0.9\textwidth,keepaspectratio]{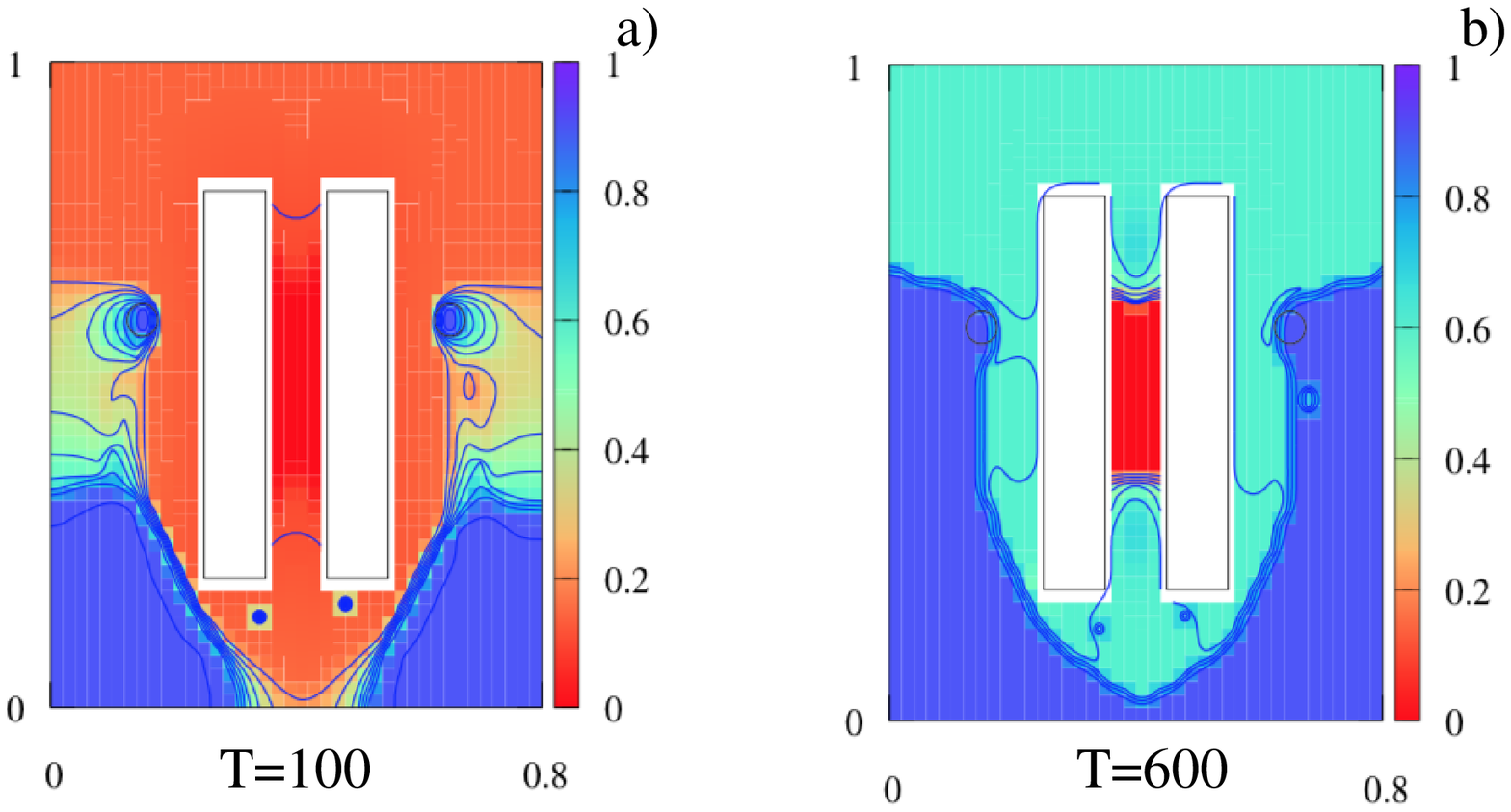}
\caption{(Case 2) spatial plots of the degree of malfunctioning  of the brain for $\alpha=10$, $\beta=1$ and $\overline{U}=0.01$ with $\tau$ at two times: a) $T=100$, b) $T=600$.}
\label{Figure2}
\includegraphics[width=0.9\textwidth,keepaspectratio]{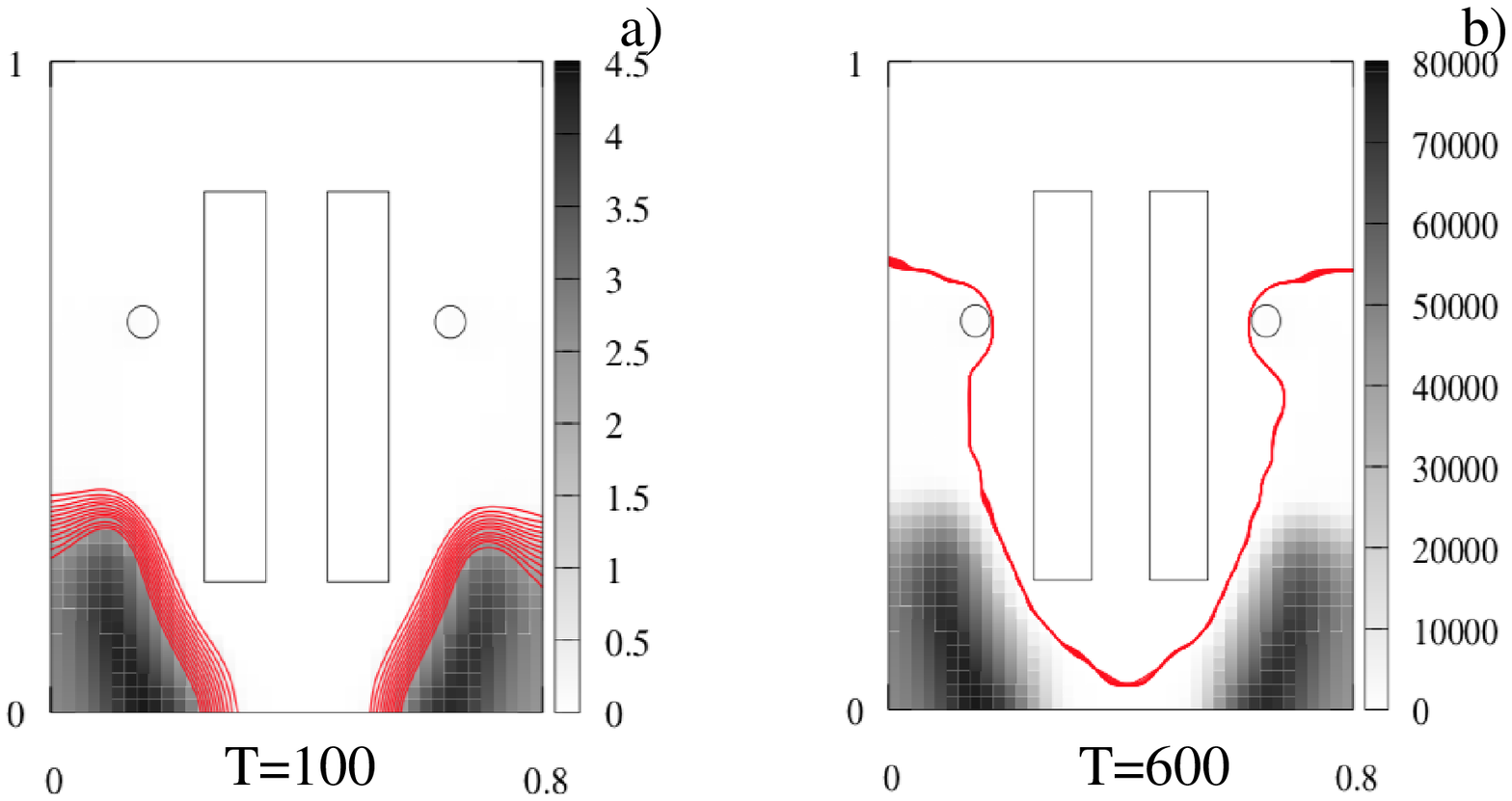}
\caption{(Case 2) spatial plots of $\tau$ for $\alpha=10$, $\beta=1$ and $\overline{U}=0.01$ at two time instants: a) $T=100$, b) $T=600$.}
\label{Figure3}
\end{figure}

Compared to~\cite{BFMTT}, the major novelty of the present paper is the inclusion of the toxic effects of  $\tau$ in the model. In the mathematical model we have added an equation describing the diffusion of $\tau$ and we have included its toxic effects on the degree of malfunctioning of the brain, as described in Section~\ref{modeltau}. In this subsection we keep the same parameter values as in Case 1 (see Table~\ref{table2}): $\alpha=10$, $\beta=1$ and $\overline{U}=0.01$.  In Figure~\ref{Figure2} we present spatial plots of the degree of malfunctioning of the brain. Numerical simulations show that  again a  numerical steady state (a quasi-steady state) is reached, which is indicated in Figure~\ref{Figure2} (b).

In this case (Case 2) the dynamics is different from that without $\tau$ (Case 1): it is faster than that  in Figures~\ref{Figure1} and follows the spreading trend of the $\tau$ concentration (compare Figures~\ref{Figure2} and ~\ref{Figure3}). Also the spatial localization of completely ill regions (the blue regions) is larger than in the case without $\tau$ (compare Figures~\ref{Figure2} and~\ref{Figure1}). These results indicate that the presence of $\tau$ influences the evolution of AD. Notice that Figure~\ref{Figure2}a is coherent with medical imaging (see, e.g.~\cite{miller2006RR} and~\cite{BFMTT}, Figure 6). As we already stressed above, previous simulations show that the presence of $\tau$ accelerates the evolution of the disease. This seems coherent with the hypothesis that different patterns for the evolution of AD (slow/fast), known in clinical literature are possibly generated by genetic variation in patients, associated with different levels of phosphorilated $\tau$ in the CSF~\cite{thalhauser_et_al,cruchaga_et_al}.

\begin{figure}[!t]
\centering
\includegraphics[width=0.9\textwidth,keepaspectratio]{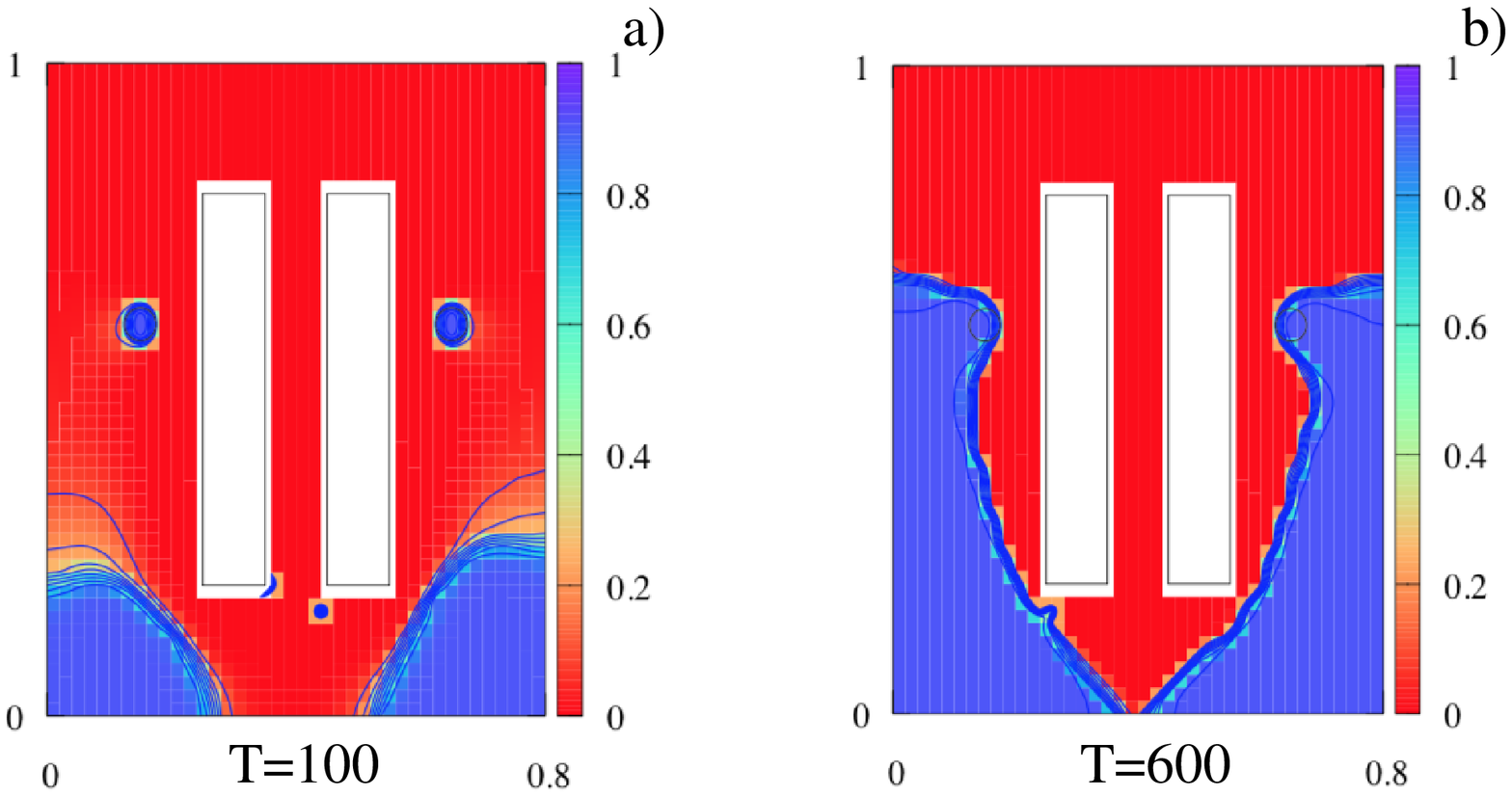}
\caption{Case 2 with $C_G=0$: 
spatial plots of the degree of malfunctioning of the brain for $\alpha=10$, $\beta=1$, $\overline{U}=0.01$ at two times: a) $T=100$, b) $T=600$.}
\label{Figure2*}
\end{figure}

Numerically one gets better insight in the role of $\tau$ by ignoring the neuron-to-neuron infection mechanism. In other words we set the constant $C_{\mathcal{G}}$ of Section~\ref{MathModel} equal to zero. Keeping the parameter values as in Cases 1 and 2, the spatial plot of the degree of malfunctioning in Figure~\ref{Figure2*} is different from that in Figure  ~\ref{Figure2}: the disease is less diffused in the cerebral parenchyma, and brain damages are localized in a portion of the brain where  $\tau$ is concentrated. Moreover, the brain is divided into two separated regions: the occipital part that is totally damaged (blue region) and the rest that is completely healthy (red region).

In conclusion, neglecting the proximity infection mechanism between neurons reduces brain damages, and this confirms the toxic effects of the synergistic interactions between A$\beta$ and $\tau$ are higher than their single toxic effects.

\subsection{The role of the agglomeration rate~\texorpdfstring{$\boldsymbol{\alpha}$}{}}
\label{sect alpha}
\begin{figure}[!t]
\centering
\includegraphics[width=0.9\textwidth,keepaspectratio]{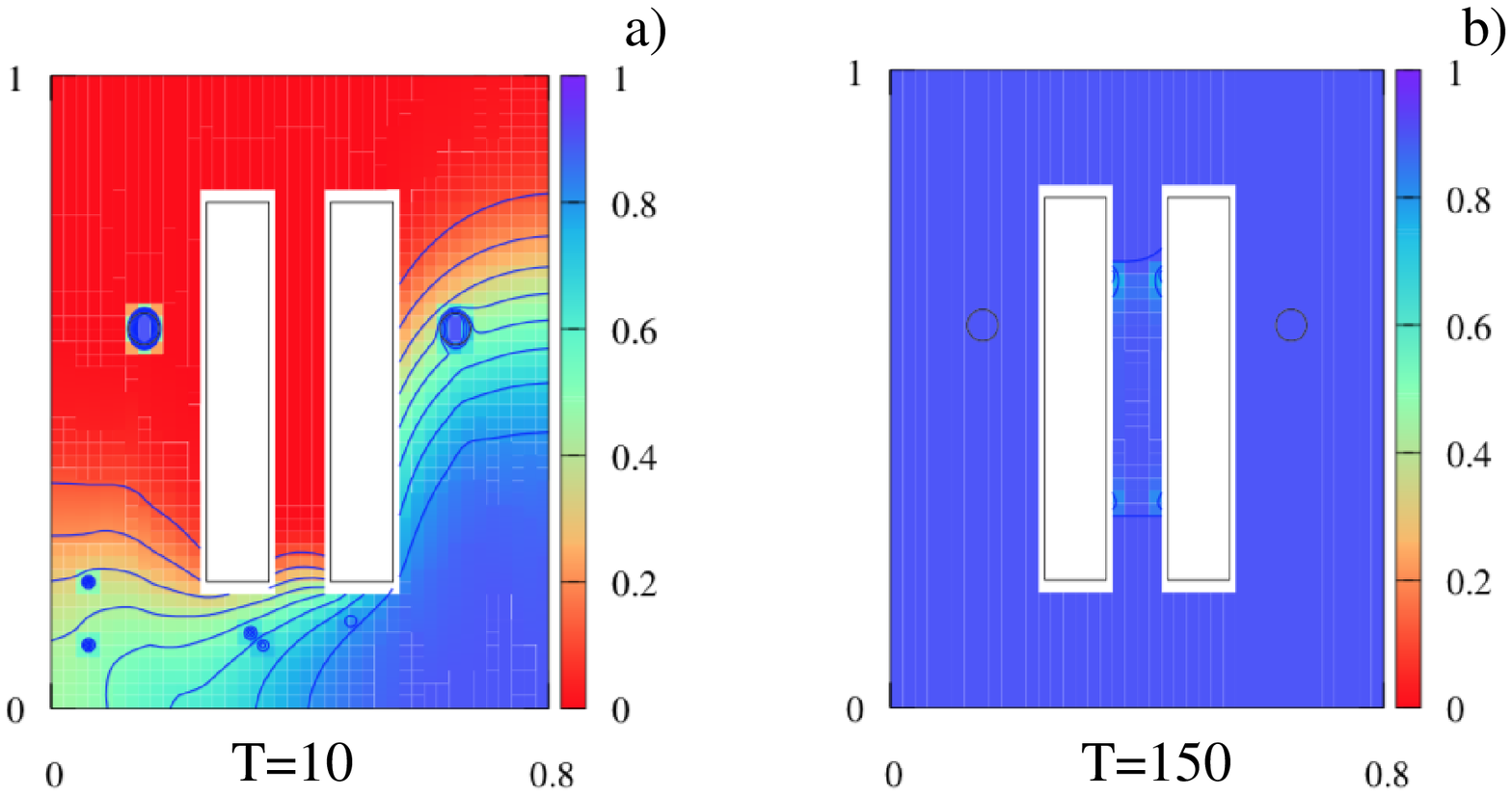}
\caption{(Case 2.1) spatial plots of the degree of malfunctioning for $\alpha=1$, $\beta=1$, $\overline{U}=0.01$ and inclusion of $\tau$ at two times: a) $T=10$, b) $T=150$.}
\label{Figure2_2}
\includegraphics[width=0.9\textwidth,keepaspectratio]{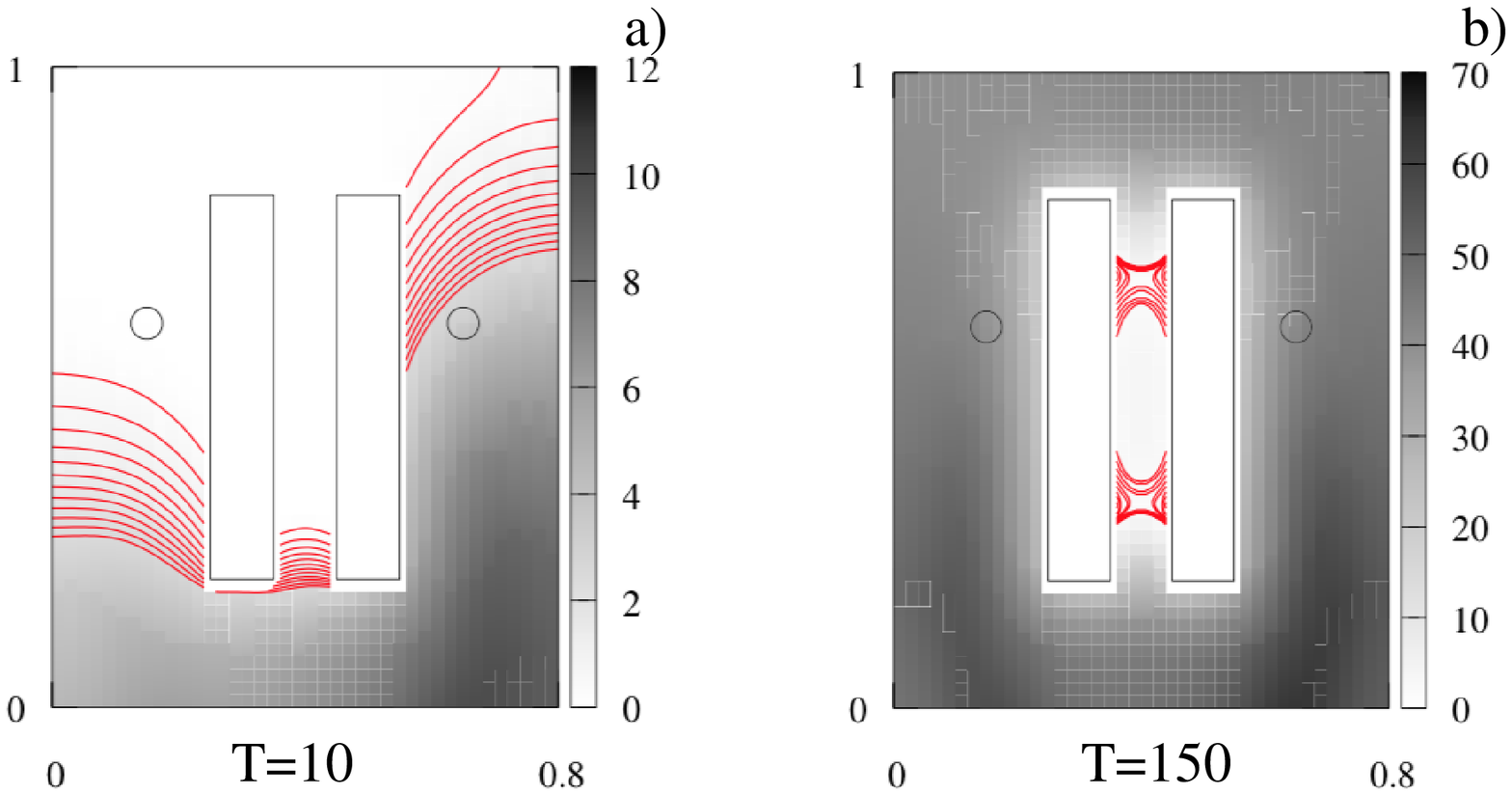}
\caption{(Case 2.1) spatial plots of $\tau$ for $\alpha=1$, $\beta=1$ and $\overline{U}=0.01$ at two time instants: a) $T=10$, b) $T=150$. }
\label{Figure2_2Tau}
\end{figure}

In this subsection we study the role of the agglomeration rate represented by the value of the parameter $\alpha$. We carry out a simulation with the parameters values of Case 2, used to obtain Figure~\ref{Figure2}, except the value of $\alpha$, which is reduced to from $10$ to $1$. This means that the aggregation of monomers is slower with respect to Figure~\ref{Figure2}. The numerical output is shown in Figure~\ref{Figure2_2}, where the spatial plots of the degree of malfunctioning of the brain are plotted at two different instants. Unlike all cases seen so far, the disease spreads faster into the whole cerebral parenchyma, until all the brain is already damaged at the computational time $T=150$. A first explanation of the phenomenon is that the transition of soluble amyloid into non-toxic plaques is much slower than in the case $\alpha=10$, and therefore A$\beta$ remains longer in the toxic oligomeric state. Moreover, looking at the spatial distribution of  $\tau$, see Figure~\ref{Figure2_2Tau}, it is clear that, unlike the previous simulations, $\tau$ spreads quickly on the whole brain, accelerating the spread of the disease and the complete damaging of the brain. From the clinical point of view, this seems to suggest that accelerating the coagulation could eventually slow down the progression of AD. However, in the current state of research, it seems unlikely to have any real potential feasibility.

\subsection{The effect of~\texorpdfstring{$\boldsymbol{\beta}$}{} in presence of~\texorpdfstring{$\boldsymbol{\tau}$}{}}
\label{sect beta}
\begin{figure}[!t]
\centering
\includegraphics[width=0.9\textwidth,keepaspectratio]{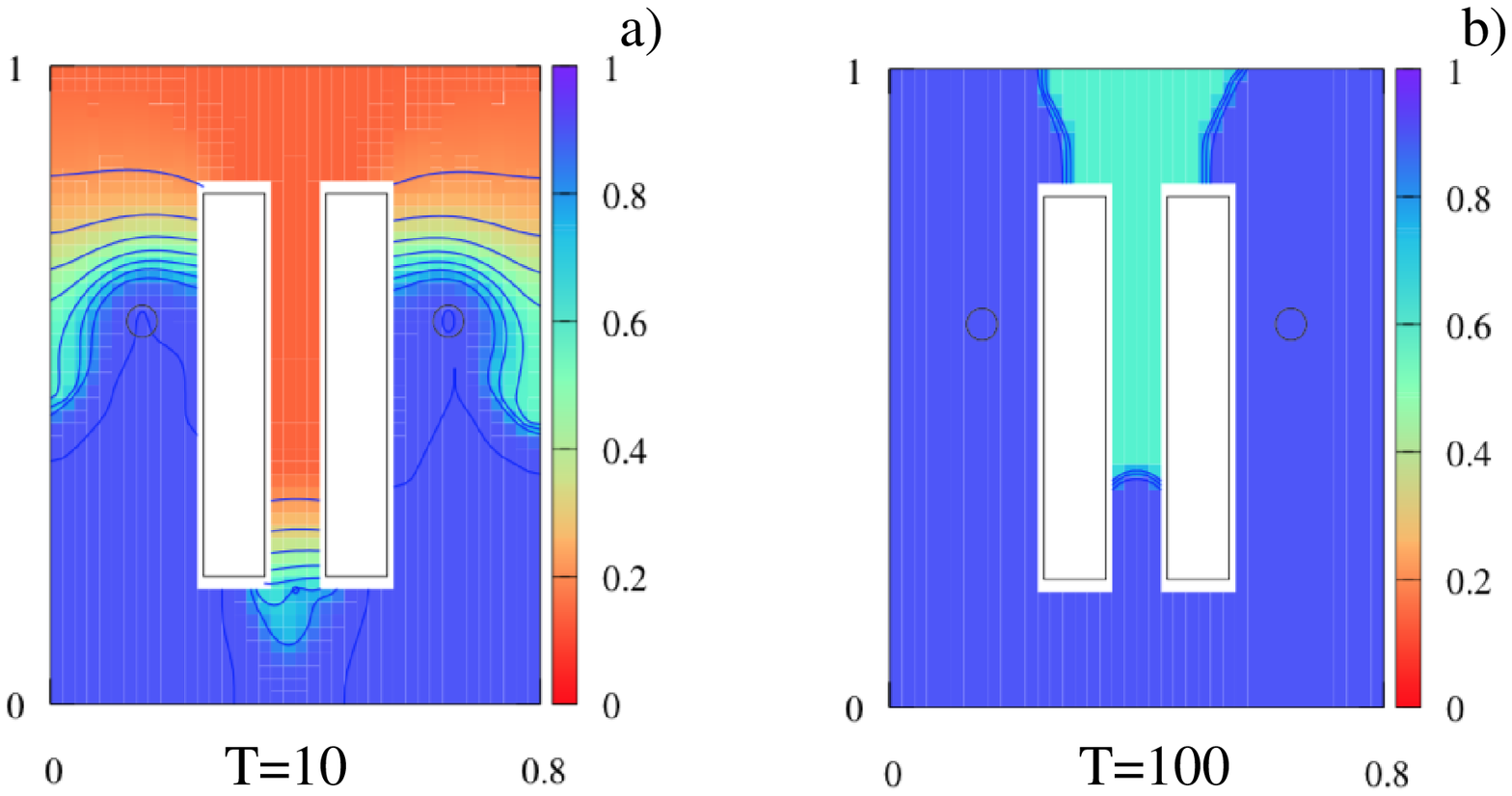}
\caption{(Case 2.2) Spatial plots of the degree of malfunctioning for $\alpha=10$, $\beta=0.01$, $\overline{U}=0.01$ and inclusion of $\tau$ at two times: a) $T=10$, b) $T=100$.}
\label{Figure2_1}
\includegraphics[width=0.9\textwidth,keepaspectratio]{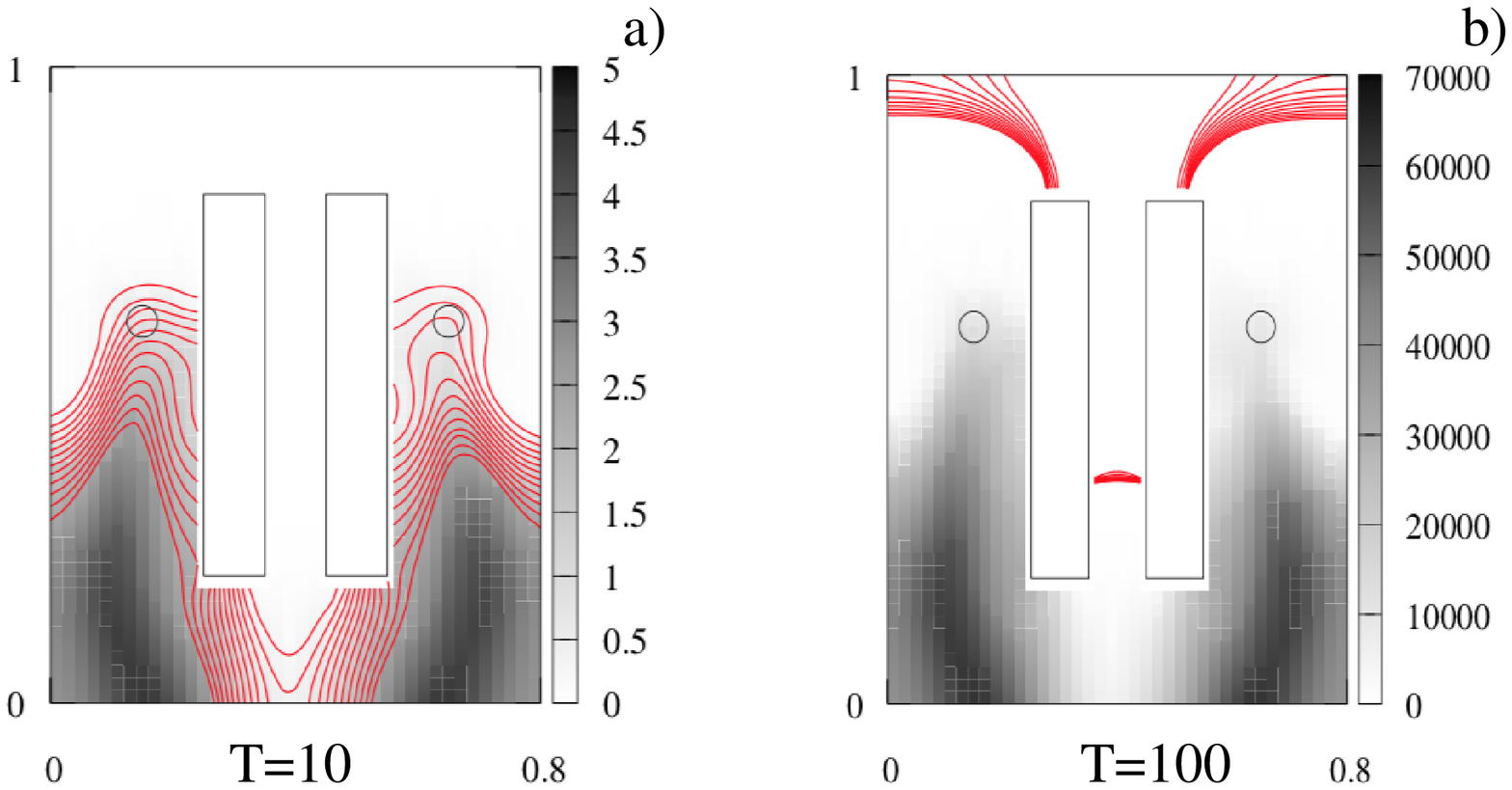}
\caption{Case 2.2) Spatial plots of $\tau$ for $\alpha=10$, $\beta=0.01$, $\overline{U}=0.01$ at two time instants: a) $T=10$, b) $T=100$.}
\label{Figure2_1Tau}
\end{figure}

This paragraph aims to investigate the effects of the parameter $\beta$, which enters the equations of the model to represent the portion of toxic A$\beta$ that is removed from the brain by clearance mechanisms  through the choroid plexus. In this simulation we use the same set of parameters values as in Figure~\ref{Figure2}, in particular keeping $\alpha=10$, except for the value of $\beta$ that we set equal to $0.01$ instead of $1$. This means that a smaller amount of toxic oligomers are removed from the brain with respect to Figure~\ref{Figure2}. In Figure~\ref{Figure2_1} some snapshots of the spatial plots of the degree of malfunctioning at two different instants are shown. The overall dynamics of the spreading of the disease is similar to that described in Figure~\ref{Figure2}, but the numerical steady state is reached faster and the completely damaged region of the brain is larger. This result appears reasonable since if less toxic elements are removed, then the degree of malfunctioning becomes higher. In agreement with what was observed in Figure~\ref{Figure2}, the spatial plots of the degree of malfunctioning reflects that of $\tau$ (see Figure~\ref{Figure2_1Tau}). Indeed, $\tau$ in the present simulation is spread in a wider region. This is not surprising because of the structure of the set of equation of our model. It might suggest to use some kind of ``dialysis'' of the CSF, but, on the other hand, this perspective seems currently unrealistic since in clinical practice it is well known that low pressure of the CSF causes headaches to the patient.

\section{Discussion}
\label{Discussion}
The brains of patients affected by Alzheimer's disease suffer from atrophy, and neuronal and synapse loss. Histopathologically they are characterized by two hallmark lesions: senile plaques containing A$\beta$ peptides, and neuro fibrillary tangles (NFT), which are composed of hyperphosphorylated forms of $\tau$-protein.

According to the Amyloid Cascade Hypothesis (ACH), A$\beta$ production and aggregation is a critical step in driving Alzheimer's disease pathogenesis. Indeed, there is substantial evidence that oligomeric A$\beta$ has a role in synapse degeneration. On the other hand, according to the ACH, A$\beta$ also triggers the hyperphosphorylation of the microtubule associated protein $\tau$ inducing circuit dysfunction.Therefore, a crucial question is how $\tau$ has to be placed in the ACH. 

Although A$\beta$ and $\tau$ show separate toxic mechanisms, their interaction is not well understood. In~\cite{ittner_et_al} the authors envisage three possible modes of interaction:
\begin{enumerate*}[label=\alph*)]
\item A$\beta$ drives $\tau$ pathology by causing hyperphosphorylation of $\tau$, which in turn mediates toxicity in neurons;
\item $\tau$ mediates A$\beta$ toxicity and hence, A$\beta$ toxicity is critically dependent on the presence of $\tau$ for example, in the dendrite;
\item A$\beta$ and $\tau$ target cellular processes or organelles synergistically, thereby possibly amplifying each others toxic effects.
\end{enumerate*}

In the present paper we developed a set of equations describing the interaction between A$\beta$ and $\tau$, assuming hypothesis a) above, however letting the model sufficiently adaptable to cover also different biological assumptions.

Our set of equations consists of a system of Smoluchowki type equations with diffusion (describing the A$\beta$ coagulation and diffusion), a transport equation describing the evolution of the disease, and an evolution equation for the hyperphosphorylation of $\tau$. We stress that our equations contains several numerical parameters that we have to tune carefully since most of them fail to have a precise value suggested by clinical data. Parameter tuning naturally leads to a sensitivity analysis of the effects of some of the relevant parameters in the model on the global dynamics of the disease.

The first insight we extrapolate from our analysis concerns the coagulation parameter $\alpha$ standing for the velocity at which monomers and oligomers aggregate. We find out that setting $\alpha$ at least equal to $10$ (see Figure~\ref{Figure2}) leads to a quasi steady state of the system in which only a portion of the brain is damaged. The location and size of the damaged portion also depend on the values of the remaining parameters of the model. This result has hypothetical clinical implications: it suggests the possibility of a drug therapy aiming to increase the aggregation rate of monomers and oligomers in order to control and localize the deterioration of the white matter. 

Concerning the effect of the parameter $\beta$, the toxic A$\beta$ removal rate, numerical simulations validate the model, showing that higher values of $\beta$ correspond to more toxic A$\beta$ removed from the cerebral parenchyma thus yielding a wider healthy region of the brain (see Figure~\ref{Figure2}) with respect to the results obtained with lower values of $\beta$ (see Figure~\ref{Figure2_1}). This is in agreement with the clinical evidence that an efficient clearance mechanism can control  the brain damages produced by the combined effect of toxic A$\beta$ and $\tau$. Finally, comparing the cases with and without $\tau$ and  keeping all other parameters fixed, we clearly observe that the evolution of the disease is much faster and the damaged brain region larger when we include $\tau$ in the model. Looking at the the spatial plots of the degree of malfunctioning, Figure~\ref{Figure2} with inclusion of $\tau$ and Figure~\ref{Figure1} without, this fact is evident. In conclusion, $\tau$ interacts with A$\beta$ accelerating the pathological dynamics of AD.  This is coherent with the hypothesis that some authors formulate to explain different patters (slow/fast) of the evolution of AD in different patients: due to unknown genetic reasons, in some individuals the CSF presents a higher rate of $\tau$ phosphorylation, thus accelerating the progress of the disease.

In addition, numerical simulations suggest a further facet of the interaction of A$\beta$ and $\tau$: if A$\beta$ can phosphorylate enough $\tau$, then we see a significant deterioration due to $\tau$. Conversely, if it cannot, then we see very little results due to $\tau$. This implies that the $\tau$ and A$\beta$ interaction may play a major role in the disease as the additional phosphorylation of $\tau$ by A$\beta$ could drive the pathology depending on its nature.

We believe that the above described numerical results and the sensitivity analysis of some of the more significant parameters validate the simplified model which we have presented. We therefore see the model as a good starting point for the development of more realistic models and their simulation in 3D. Among other things, we mention that the complexity of the brain should be taken into account, including an anatomically correct picture of the neural bundles (that also give rise to anisotropic diffusion), and that the CFS clearing system should be modelled, including the possible role of the so-called glymphatic system. The latter topic is particularly challenging and complex, and obviously cannot be described by a single parameter such as our $\beta$.

\section*{Acknowledgements}
The authors express their deepest gratitude to the referee: his extensive and precise review was for them a source of
most valuable suggestions, comments, scientific facts and references, which they used to prepare a revised and improved version of the paper.    

\medskip

The authors would like to express their gratitude to MD Norina Marcello for many stimulating and fruitful discussions over several years.

B.F. and M.C.T. are supported by the University of Bologna, funds for selected research topics, by MAnET Marie Curie Initial Training Network and by GNAMPA of INdAM (Istituto Nazionale di Alta Matematica ``F. Severi''), Italy.

M.B. thanks the project Beyond Borders (CUP E84I19002220005) of the University of Rome Tor Vergata.
M.B. and V.M. acknowledge the MIUR Excellence Department Project awarded to the Department of Mathematics, University of Rome Tor Vergata, CUP E83C18000100006.

A.T. acknowledges support by the Italian Ministry of Education, University and Research (MIUR) through the ``Dipartimenti di Eccellenza'' Programme (2018-2022) -- Department of Mathematical Sciences ``G. L. Lagrange'', Politecnico di Torino (CUP:E11G18000350001) and through the PRIN 2017 project (No. 2017KKJP4X) ``Innovative numerical methods for evolutionary partial differential equations and applications''. This work is also part of the activities of the Starting Grant ``Attracting Excellent Professors'' funded by ``Compagnia di San Paolo'' (Torino) and promoted by Politecnico di Torino. Moreover, A.T. acknowledges membership of GNFM (Gruppo Nazionale per la Fisica Matematica) of INdAM (Istituto Nazionale di Alta Matematica), Italy.


\end{document}